\documentclass[aps,prl,floatfix,epsfig,twocolumn,showpacs,preprintnumbers]{revtex4}
\usepackage{amsmath}
\usepackage{graphicx}
\usepackage{epstopdf}
\usepackage{dcolumn}
\usepackage{bm}
\usepackage{mathrsfs}
\usepackage[pdfstartview=FitH, CJKbookmarks=true, bookmarksnumbered=true, bookmarksopen=true, pdfborderstyle={/S/U}, pdfborder={0 0 1}, colorlinks=true, linkcolor=blue, anchorcolor=green, citecolor=blue]{hyperref}

\begin{document}

\title{Correlation Driven Topological Insulator--to--Weyl Semimetal
Transition in Actinide System UNiSn}
\author{Vsevolod Ivanov$^{\dag }$, Xiangang Wan$^{\ast }$, Sergey Y. Savrasov%
$^{\dag }$}
\affiliation{$^{\dag }$Department of Physics, University of California, Davis, CA 95616,
USA}
\affiliation{$^{\ast }$Department of Physics, Nanjing University, Nanjing 210093, China}

\begin{abstract}
Although strong electronic correlations are known to be responsible for some
highly unusual behaviors of solids such as, metal--insulator transitions,
magnetism and even high--temperature superconductivity, their interplay with
recently discovered topological states of matter awaits full exploration.
Here we use a modern electronic structure method combining density
functional theory of band electrons with dynamical self--energies of
strongly correlated states to predict that two well--known phases of
actinide compound UNiSn, a paramagnetic semiconducting and antiferromagnetic
metallic, correspond to Topological Insulator\ (TI) and Weyl semimetal (WSM)
phases of topological quantum matter. Thus, the famous unconventional
insulator--metal transition observed in UNiSn is also a TI--to--WSM
transition. Driven by a strong hybridization between U f-electron multiplet
transitions and band electrons, multiple energy gaps open up in the
single--particle spectrum whose topological physics is revealed using the
calculation of $Z_{2}$ invariants in the strongly correlated regime. A
simplified physical picture of these phenomena is provided based on a
periodic Anderson model of strong correlations and multiple band inversions
that occur in this fascinating compound. Studying the topology of
interacting electrons reveals interesting opportunities for finding new
exotic phase transitions in strongly correlated systems.
\end{abstract}

\date{\today }
\maketitle

\begin{figure}[tbp]
\includegraphics[height=0.235\textwidth,width=0.45%
\textwidth]{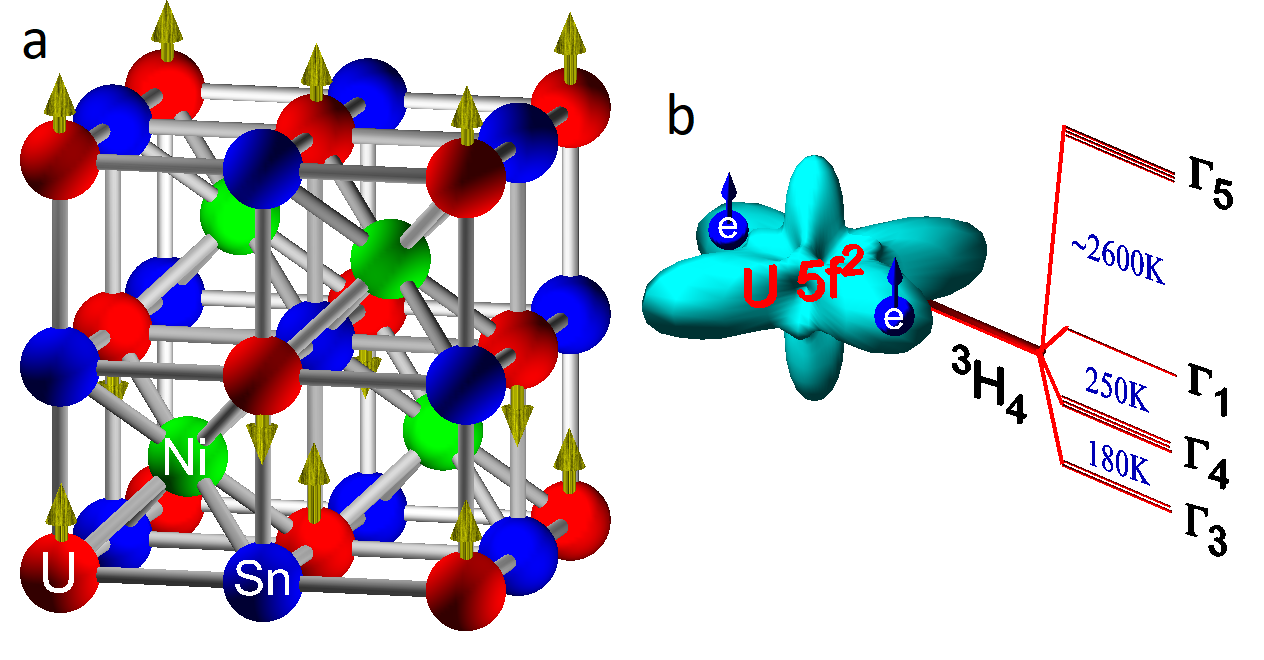}
\caption{a. Crystal structure of UNiSn showing antiferromagnetic type I
ordering \protect\cite{UNiSn}, b. Effect of the cubic crystal field
splitting on the $^{3}$H$_{4}$ ground state multiplet of the U f$^{2}$
two--electron state with its lowest non--magnetic $\Gamma _{3}$ doublet as
found experimentally\protect\cite{Aoki}.}
\label{FigStruc}
\end{figure}

Here we argue that an unusual phase transition at T$_{N}$=43K between a
higher temperature paramagnetic semiconducting (P--S) and low temperature
antiferromagnetic metallic (AFM)\ phase discovered in the past \cite{UNiSn}
for a strongly correlated actinide system UNiSn, simultaneously corresponds
to the transition between two topological phases of quantum matter,
Topological Insulator (TI) and Weyl semimetal (WSM), that have recently
received a great interest due to the appearance of robust electronic states
insensitive to perturbations\cite{RMPTI, RMPWSM}. This actinide compound has
been extensively studied during last several decades owing to its
unconventional\ (inverse) nature of this metal--insulator transition with
the gap opening above T$_{N}$ and associated behavior of its strongly
correlated 5f electrons. It crystallizes in a cubic structure (MgAgAs--type)
(see Fig. \ref{FigStruc}a) and its paramagnetic semiconducting phase has an
estimated energy gap of about 100 meV \cite{Riseborough}. Its
antierromagnetic structure was found to be of type I with the ordered U
moment 1.55 $%
\mu
_{B}$ oriented along the (001) axis \cite{UNiSn}. 

The central issue in understanding the physical properties of actinides is
the degree to which their 5f electrons are localized. Due to the absence of
any signatures of heavy fermion behavior in the specific heat data\cite%
{Riseborough}, the magnetic behavior of UNiSn has been explained \cite{Aoki}
on the basis of a localized 5f$^{2}$ (U$^{4+}$) ionic state, whose ground
state multiplet $^{3}H_{4}$ $(J=4)$ subjected to a cubic crystal field is
split into a doublet ($\Gamma _{3}),$ two triplets ($\Gamma _{4},\Gamma
_{5}) $ and a singlet $(\Gamma _{1})$. Measured temperature--dependent
susceptibility and magnetic entropy analysis suggested that the
non--magnetic doublet is the lowest lying state 180K below the $\Gamma _{4}$
triplet and 430 K below the $\Gamma _{1}$singlet (see Fig. \ref{FigStruc}b).
Since $\Gamma _{3}$ has a quadrupole moment, it was further proposed that
tetragonal distortions and quadrupolar ordering exists below T$_{N}$\cite%
{UNiSn}. The valence band photoemission spectra revealed a dominant 5f
electron character for the states in the vicinity of the Fermi level with a
contribution from U 6d, Ni 3d and Sn 6p states \cite{PES}.

Previous band structure calculations of UNiSn emphasized the role of
relativistic effects and electronic correlations among 5f electrons \cite%
{UNiSn-LDA+U}. Both P--S and AFM behavior have been captured correctly
within the LDA+U framework\cite{LDA+U}, where on--site Coulomb correlations
among f electrons are treated via the introduction of the Hubbard U term and
subsequent static mean field approximation. Such a method is expected to
work well in a symmetry broken AFM\ state, but would be invalid for the
genuine two--electron $\Gamma _{3}$ doublet represented by a mixture of
Slater determinants. One can however, assume that paramagnetism originates
from the non--magnetic $\Gamma _{1}$ singlet, for which LDA+U should be
sufficient. Within a single--particle picture this is interpreted as a
doubly occupied $\Gamma _{7}$ level that appears when a 14 fold degenerate
manifold of 5f electrons subjected to spin--orbit coupling and cubic crystal
field is split into $\Gamma _{7},\Gamma _{8}$ (for j=5/2) and $\Gamma
_{6},\Gamma _{7},\Gamma _{8}$ (for j=7/2) sublevels. Detailed comparisons
between theory and experiment revealed discrepancies in the position of the
occupied f--band with respect to the Fermi energy: --0.3 eV in the
photoemission vs. --1 eV in the LDA+U calculation\cite{PES}.

Nowadays, modern electronic structure approaches based on combinations of
local density approximation and dynamical mean field theory (so called
LDA+DMFT) \cite{LDA+DMFT} are free from these difficulties, and allow for a
more accurate treatment of Coulomb interactions\ in UNiSn via computations
of local self--energies $\Sigma (\omega )$ for the interacting 5f electrons.
This is achieved by treating a correlated 5f shell as an impurity hybridized
with the non--interacting bath which is then periodized\ and subjected to
self--consistency. An exact solution of such an Anderson impurity problem is
possible in principle via a recently developed Continuous Time Quantum Monte
Carlo (CT--QMC) method \cite{CTQMC}, although accounting for the full
Hilbert space of interacting f electrons together with spin--orbit and
crystal field terms represents a challenge. In addition, the CT--QMC works
on the imaginary time--frequency axis and obtaining frequency dependence of
the self--energy on the real axis involves an analytical continuation
algorithm which is known to be not very accurate.

\begin{figure}[tbp]
\includegraphics[height=0.475\textwidth,width=0.40%
\textwidth]{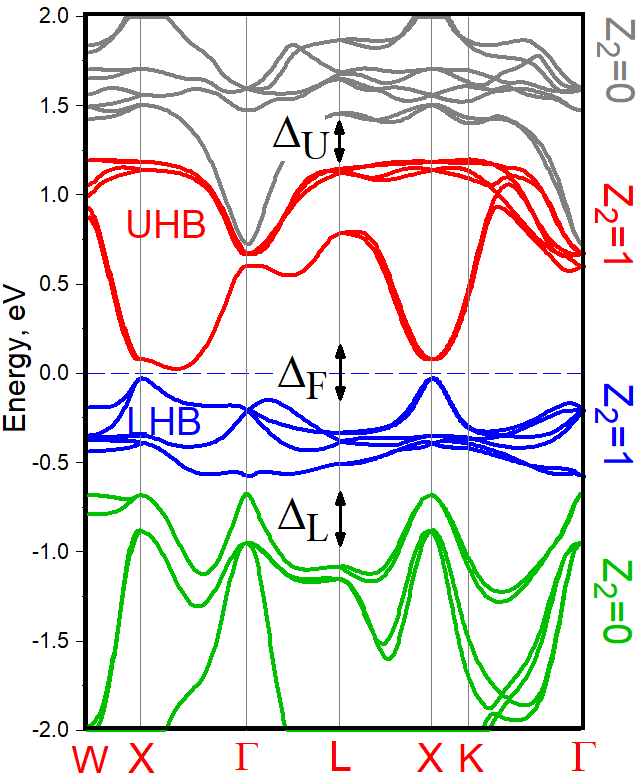}
\caption{Calculated electronic structure of UNiSn using density functional
theory combined with dynamical self--energies for the Uranium f--electrons
assuming experimentally determined 5f$^{2}$ $\Gamma _{3}$ doublet as a
ground state\protect\cite{Aoki}, The locations of energy panels with
non--zero $Z_{2}$ invariants and corresponding gaps ($\Delta _{L},\Delta
_{F},\Delta _{U}$) are indicated. }
\label{LDA+DMFT-1cell}
\end{figure}

In order to study the topology of correlated electrons in UNiSn here we take
a pragmatic approach and make the problem numerically tractable by using the
experimental fact that the Uranium f electrons are localized in their 5f$%
^{2} $ $\Gamma _{3}$ ground state, from which the one--electron multiplet
transitions can be obtained by exact diagonalization. The corresponding
f--electron self--energies are subsequently expanded in the Laurent series
which allows us to replace the non--linear (in energy) Dyson equation by a
linear Schroedinger--like equation in an extended subset of
\textquotedblleft pole states\textquotedblright\ \cite{Am}. Remarkably, the
pole representation for the self--energy results in the appearance of
many--body satellites and multiplets in the spectra as effective band
states, in general carrying a fractional occupancy due to the spectral
weight transfer. It is ideally suited for studying topological indices as
the corresponding auxiliary wave functions representing the many--body
features carry all necessary information about the Berry phase of the
interacting electrons\cite{XiDai}.

We now present the results of our\ calculation for the paramagnetic phase of
UNiSn which is carried out by treating f--electrons in their 5f$^{2}$ $%
\Gamma _{3}$ ground state. The Coulomb interaction matrix elements\ needed
for the exact diagonalization procedure ($F^{(0)},F^{(2)},F^{(4)},F^{(6)}$
Slater integrals) have been found from the atomic 5f--electron wave
functions and scaled to account for screening effects. We cover a range of
these parameters: 2--4 eV for the Hubbard $U=F^{(0)}$ and 0--1 eV for the
exchange $J=(286F^{(2)}+195F^{(4)}+250F^{(6)})/6435$, in order to make sure
that our conclusions are not altered by the lack of an accurate procedure
for determining the screening. It has been argued earlier that these values
are typical for obtaining the best agreement between theory and experiment
for several Uranium compounds \cite{PES,UNiSn-LDA+U}. The position of the
bare f--level is fixed by reproducing the experimentally observed $%
f^{2}\rightarrow f^{1}$ electron removal transition at --0.3 eV \cite{PES}.
The charge density self--consistency is carried out within the LDA+DMFT as
implemented by one us earlier\cite{PuNature}. 
\begin{figure*}[tbh]
\includegraphics[height=0.417\textwidth,width=0.95%
\textwidth]{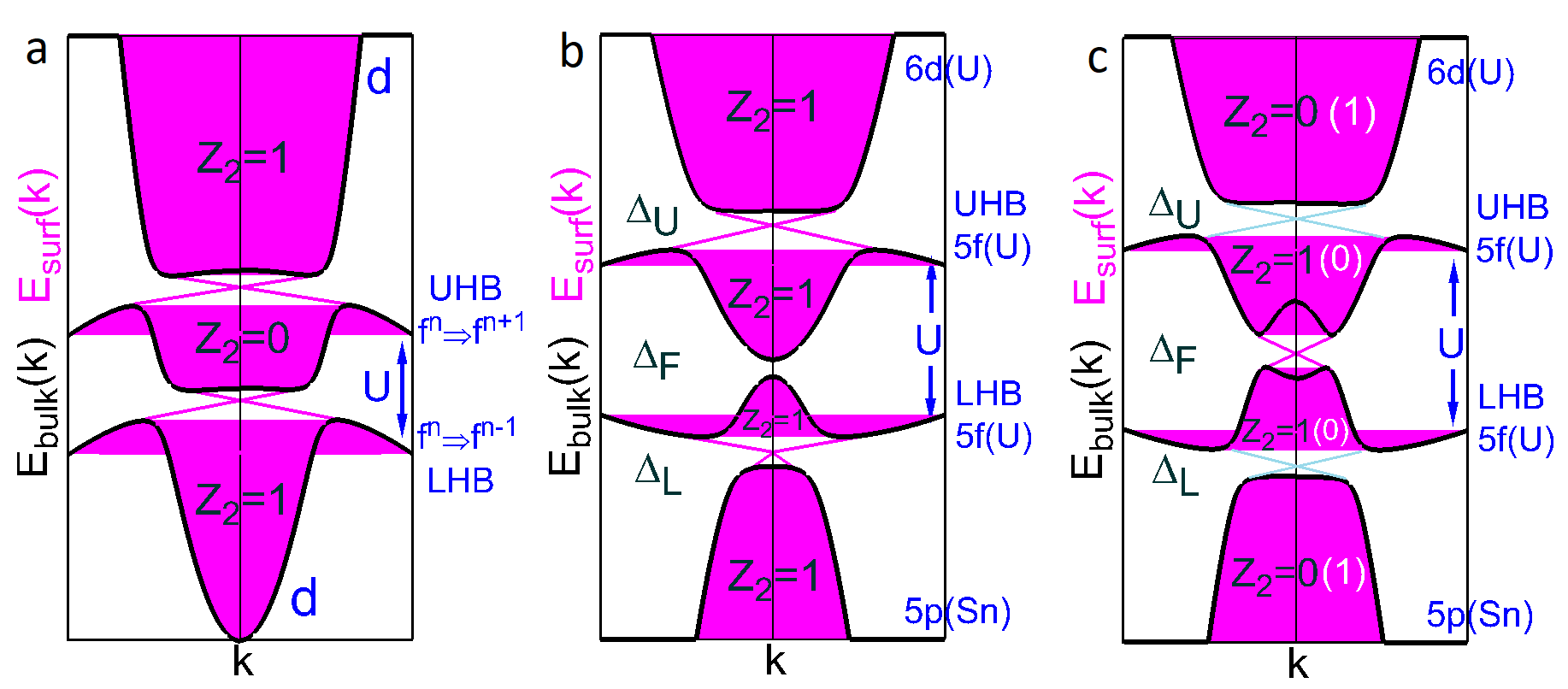}
\caption{Band inversion mechanism applicable for UNiSn: a. In the periodic
Anderson model, hybridization between a wide d-band and f--electron
multiplet transitions (lower and upper Hubbard bands, LHB and UHB) results
in three energy panels (shown by black lines) and two gaps that are both
topologically non-trivial. The corresponding surface spectrum is shown by
magenta color, where the spectral weight\ of the Dirac cone is distributed
between the two gaps. b. In UNiSn, the upper Hubbard band is inverted with
the wide 6d band of Uranium while the lower Hubbard band is inverted with
the 5p band of Tin, resulting in four energy panels and three gaps. With two
such band inversions, upper, $\Delta _{U},$ and lower, $\Delta _{L},$ the
hybridization gaps are topological while the fundamental bulk gap, $\Delta
_{F},$ is not. c. The band inversion between U 6d and Sn 5p states makes the
fundamental gap $\Delta _{F}$ topological. The topological features of the
gaps $\Delta _{U},\Delta _{L}$ are seen to disappear in the LDA+DMFT
calculation, but this is not a requirement within the considered model ($%
Z_{2}$ invariants shown in parentheses are expected).}
\label{UNiSnModel}
\end{figure*}

Fig. \ref{LDA+DMFT-1cell} shows our calculated many--body electronic
spectrum in the vicinity of the Fermi level using a set of Slater integrals $%
F^{(0)}=0.15,F^{(2)}=0.3,F^{(4)}=0.2,F^{(6)}=0.15$ in Rydberg units.\
Although cast into a conventional band structure plot, we stress that the 5f
electron states are treated here as true one--electron removal (f$%
^{2}\rightarrow $f$^{1}$) and addition (f$^{2}\rightarrow $f$^{3}$)
processes that come from exact diagonalization, and corresponding "energy
bands" carry non--integer occupation. This can be seen by realizing that the
multiplet transitions within the $j=5/2$ manifold (shown in this Figure by
red and blue) are represented by 6 energy bands that appear both below and
above the Fermi level. These are the famous lower and upper Hubbard bands
within the Mott gap picture that acquire a significant dispersion due to
hybridization with U 6d and Sn 5p orbitals. The deduced value of the
indirect energy gap shows some dependence on the Slater integrals, but falls
into the same range as experiment (\symbol{126}100 meV \cite{Riseborough}).

\begin{figure}[tbp]
\includegraphics[height=0.313\textwidth,width=0.45%
\textwidth]{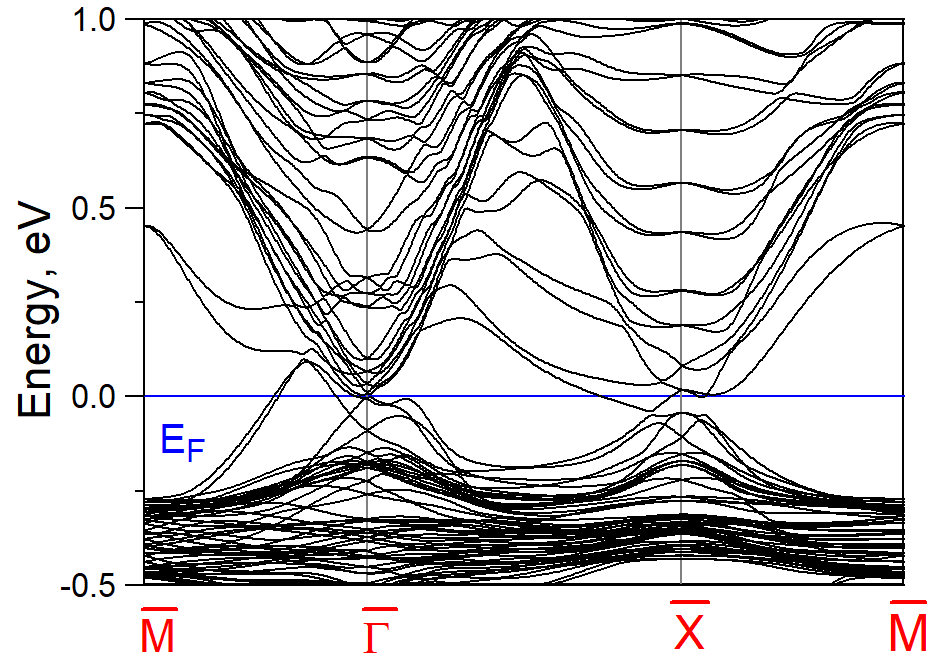}
\caption{Surface spectrum of UNiSn calculated using LDA+DMFT method,
assuming a 5f$^{2}$ $\Gamma _{3}$ doublet as a ground state. The fundamental
energy gap shows topological Dirac cone dispersions in the vicinity of the $%
\bar{\Gamma}$ and $\bar{X}$ points. }
\label{LDA+DMFT-surf}
\end{figure}

We now turn to the prediction of topological properties of the paramagnetic
semiconducting phase of UNiSn. First, we point out that the underlying
crystal structure is not centro--symmetric, therefore the Fu and Kane parity
criterion\cite{FuKane} developed for insulators with both time reversal and
inversion symmetries does not apply. Nevertheless, given the fact that the
Uranium sites arrange themselves in the inversion symmetric face centered
cubic sublattice with their odd parity localized 5f electrons lying in close
proximity to the Fermi level, it is interesting to speculate whether the
possibility of inversion with the even parity U 6d band is taking place.
Such an f--d band inversion was at the center of recent interest for several
topological Kondo insulator materials with 4f electrons\cite{TKI}, as well
as in some actinide systems such as AmC\cite{AmC}. While the U 6d band is
expected to be unoccupied, it is very wide, with its lower portion
hybridized with the Hubbard bands. The Fu and Kane criterion would then
imply the existence of topological Dirac cone states in UNiSn.

\begin{figure*}[tbh]
\includegraphics[height=0.264\textwidth,width=0.95%
\textwidth]{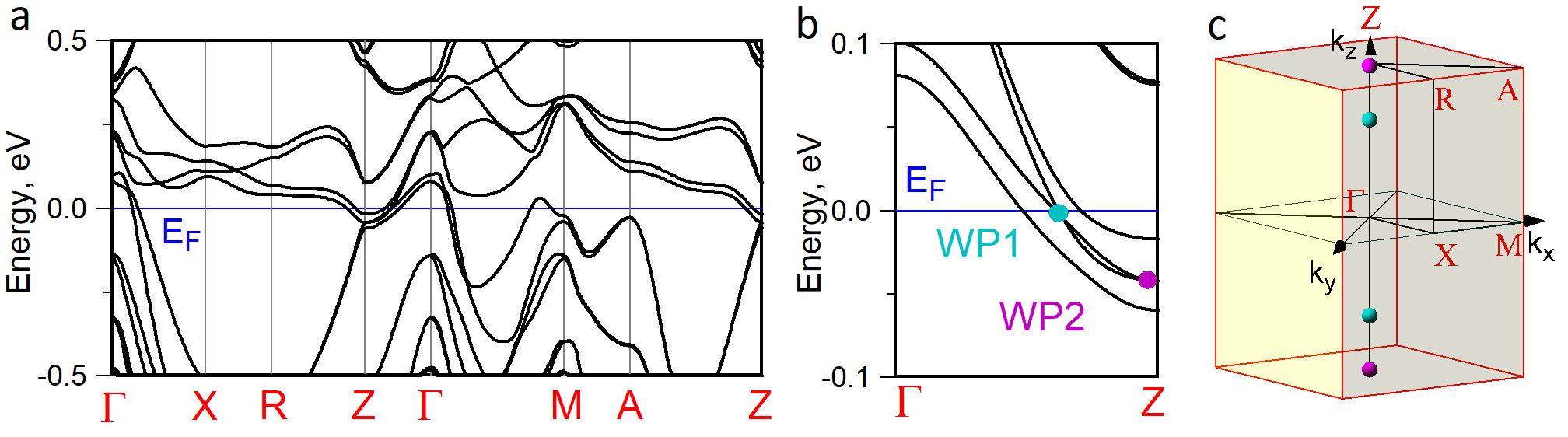}
\caption{Calculations for antiferromagnetic configuration of UNiSn. a.
Energy band dispersions along major high symmetry lines of the Brillouin
Zone, b. Zoomed area along the $\Gamma Z$ direction of the BZ showing the
locations of the Weyl points, c. Brillouin Zone of the AFM\ UNiSn with the
positions of the Weyl points.}
\label{LDA+DMFT-AFM}
\end{figure*}

To uncover the topological physics one needs to compute $Z_{2}$ invariants 
\cite{Z2} for the occupied band manifold in the difficult regime of strong
correlations. Fortunately, it was recently proved that utilizing a pole
representation for the self--energy\cite{Am}, reduces this problem to an
effective non--interacting system in the extended set of pole states, whose
topological indices are exactly matched \cite{XiDai}. We develop and carry
out this computation within the n--field approach\cite{n-field}. However,
some care should be taken to define an appropriate energy panel because as
it is seen from our calculations that multiple gaps appear in the
excitational spectrum of UNiSn (we show the panels by various colors and
denote the gaps between them as $\Delta _{L},\Delta _{F},\Delta _{U}$ in
Fig. \ref{LDA+DMFT-1cell}). For example, six dispersive features that
represent the lower Hubbard bands (blue colored "spaghetti" in Fig. \ref%
{LDA+DMFT-1cell} labeled as LHB) are completely gapped from the remaining
band manifold everywhere in the BZ. The same is seen for the six eigenstates
representing the upper Hubbard bands above the E$_{F}$ (red colored
"spaghetti" in Fig. \ref{LDA+DMFT-1cell} labeled UHB)$.$ Our computations of
Z$_{2}$ invariants for the four energy panels separated by $\Delta
_{L},\Delta _{F},\Delta _{U}$ reveal their topological indices that we
indicate on the right margin in Fig. \ref{LDA+DMFT-1cell}. The energy panels
below and above the fundamental gap correspond to the indices equal to
1;(000) in the notations of Ref. \cite{Z2} \ (we denote this result by $%
Z_{2}=1$ in Fig. \ref{LDA+DMFT-1cell}). This suggests that UNiSn is a strong
topological insulator and prompts on the existence of protected Dirac cone
states at its surface.

To understand which orbitals are responsible for the appearance of the
topological phase, we carry out calculations using a constrained
hybridization approach\cite{XGW}. In this method, the energies of particular
orbitals are shifted by applying a constant potential constrained within the
orbital space by projector operators. This is similar to the LDA+U, LDA+DMFT
and other families of methods that combine the self--energy with LDA\
(SELDA\ family \cite{SELDA}), restricting the application of the
self--energy to the subspace of correlated orbitals. Utilizing this
procedure, we are able to de--hybridize various states, such as\ U-5f, U-6d,
Ni-3d, Sn-5p, etc., by shifting their energies away from the relevant energy
window, and recompute Z$_{2}$ invariants. The outcome of this study is the
existence of multiple band inversions in UNiSn: The upper Hubbard band is
inverted with the U 6d electrons, the lower Hubbard band is inverted with Sn
5p electrons, and, most importantly, U 6d electrons at the\ very bottom of
the conduction band and Sn 5p states at the very top of the valence band are
also inverted around the X point of the BZ (see Fig. \ref{LDA+DMFT-1cell}).
This band inversion is responsible for the topological insulator property of
UNiSn.

To illustrate the emergent physical picture, we use the periodic Anderson
model (PAM) of strong correlations. It has been recently employed for
developing the concept of topological Kondo insulators where the Fermi level
falls into the gap between a heavy fermion (f--like) and non--interacting
(d--like) bands\cite{TKI}. It has also been recently used to describe
Weyl--Kondo semimetals via hybridization of a heavy--fermion state with
non--interacting bands containing the nodal points\cite{WeylKondo}. In our
case the f--electrons are localized and their self--energies behave
similarly to the famous Hubbard I approximation: $\Sigma (\omega
)=U^{2}/4\omega .$ The solution of the PAM in this limit is schematically
illustrated in Fig. \ref{UNiSnModel}a. Hybridization between a wide d--band
and f--electron multiplet transitions denoted as LHB and UHB results in the
appearance of two gaps in the spectrum and three energy panels (shown by
black lines). Both gaps are seen to be topologically non--trivial due to the
d--f band inversion mechanism. For a centrosymmetric lattice, this can be
understood based on the Fu and Kane parity criterion \cite{FuKane}: for the
lower (upper) panel, the parities of the eigenstates are odd (even) at BZ
boundary time reversal invariant momenta (TRIM),\ such as the $X$ and $L$
points of cubic BZ, but even (odd) at $\Gamma .$ This results in the energy
gap above (below) the panel to be topological. For the central panel, the
parities of the eigenstates are odd everywhere, but this does not preclude
having topological gaps below and above the panel, each with its own Dirac
cone (the total number of cones is even). We illustrate the corresponding
surface spectrum by magenta color. Note that since the Hubbard bands carry
no integer occupation, the spectral weight\ of the Dirac cones is also
re--distributed between the two gaps.

Now, in UNiSn, our constrained hybridization procedure reveals multiple band
inversions: First, as illustrated in \ref{UNiSnModel}b, the upper Hubbard
band is inverted with the U 6d, while the lower Hubbard band is inverted
with Sn 5p. Due to the d--f band inversion, the topological Dirac cone is
expected to appear inside the gap $\Delta _{U}$ at the surface spectrum.
Since Uranium atoms occupy sites of the centrosymmetric face centered cubic
lattice, this can be understood based on the parity criterion \cite{FuKane}.
For the lower Hubbard band, U 5f and Sn 5p orbitals are both odd, but they
occupy different sites (forming a diamond--like lattice) where the inversion
center can be imagined approximately in the middle between the atoms. Such
model will also produce a strong topological insulator (the Dirac cone is
inside the gap $\Delta _{L})$ provided that the irreducible representations
of the U 5f and Sn 5p orbitals are different, e.g. $\Gamma _{7}$ and $\Gamma
_{6}.$ This picture emerges when the bottom of the U 6d and the top of Sn 5p
bands are not inverted around the X point, making the fundamental gap $%
\Delta _{F}$ not topological, as we show in \ref{UNiSnModel}b. Realizing the
band inversion between the U 6d and Sn 5p bands at the X point (see Fig. \ref%
{UNiSnModel}c) results in the fundamental gap $\Delta _{F}$ becoming
topological. Additionally, we monitor the cancellation of the topological
features inside the gaps $\Delta _{U},\Delta _{L}.$ This is apparently due
to a more complex overlap between various orbitals in the real calculation
than the one assumed in the simplified model illustrated in Fig. \ref%
{UNiSnModel}c where one would in principle expect all three gaps to become
topological ($Z_{2}=0$ for the LHB\ and UHB, and $Z_{2}=1$ for the lowermost
and topmost panels as shown in brakets in Fig. \ref{UNiSnModel}c).

\begin{figure}[tbp]
\includegraphics[height=0.38\textwidth,width=0.45\textwidth]{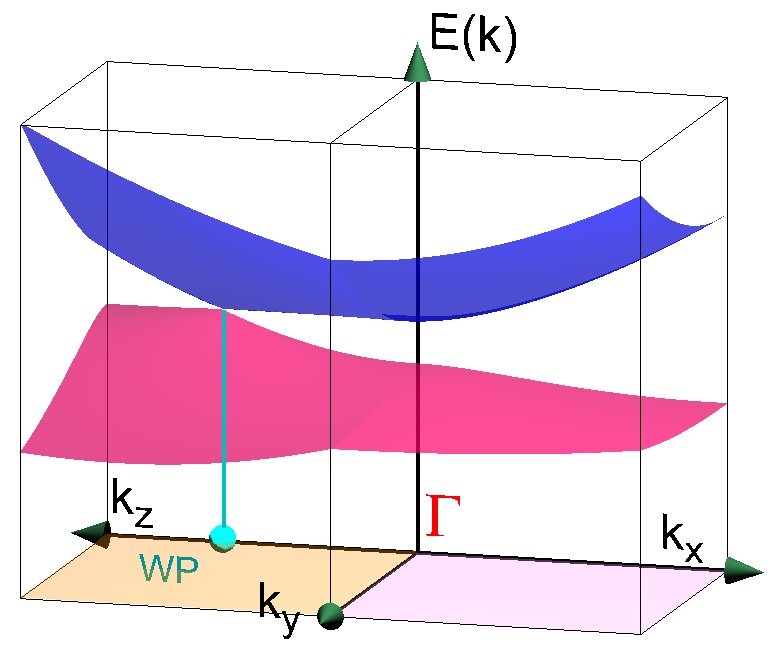}
\caption{Dispersion of two intermediate eigenstates of a 4x4 k*p model used
to illustrate the magnetization induced Weyl semimetal state in UNiSn. The
band structure is gapped for all k--points in the BZ except along the $%
\Gamma Z$ line where the Weyl point is formed. }
\label{FigKP}
\end{figure}

To verify these conclusions, we additionally carry out\ the charge density
self--consistent LDA+DMFT\ calculation of the surface one--electron
spectrum. We construct a slab spanned along z direction and terminated by U
plane from the top and by Ni plane from the bottom. The slab sizes are
varied between 4 and 8 unit cells and the distance between the slabs is set
to 14 \AA\ to ensure the convergence of the surface states. The results are
shown in Fig. \ref{LDA+DMFT-surf}. We clearly resolve the Dirac cone states
inside the fundamental energy gap $\Delta _{F}$ of UNiSn that appear around
the $\bar{\Gamma}$ of the surface BZ. There are also Dirac cone--like
features around the $\bar{X}$ point. It is interesting to note that there
are other surface states propagating across the gap and hybridizing with the
Dirac cones. This makes the Dirac cone dispersions appear within small areas
around the TRIM points, but break apart at higher momenta.

We now turn to discussing the results of our calculation for the low
temperature AFM\ phase of UNiSn. The origin of magnetism has been explained
earlier \cite{Aoki} based on a molecular--field model, where owing to the
second--order effect in the magnetic exchange field, the $\Gamma _{3}$
doublet is split into two levels with the deduced value of magnetic moment $%
\sim 2.6\mu _{B}.$ Here our exact diagonalization for the 5f states is
almost identical to to the static mean field solution, because the double
degeneracy of $\Gamma _{3}$ is broken and a single Slater determinant
description suffices. It has been also proven earlier that the LDA+DMFT
method reduces to the LDA+U in the Hartree--Fock limit\cite{Imseok}. Our
calculation with the Slater integrals $%
F^{(0)}=0.15,F^{(2)}=0.3,F^{(4)}=0.2,F^{(6)}=0.15$ in Rydberg units
converges to an antiferromagnetic state with the total magnetic moment of 2.1%
$\mu _{B}$ (+3.2$\mu _{B}$ for its orbital and -1.1$\mu _{B}$ for its spin
counterparts) slightly larger than the experimentally deduced value of 1.55 $%
\mu _{B}$ \cite{UNiSn}. This is in agreement with previous works \cite%
{PES,UNiSn-LDA+U} that also pointed out the inclusion of spin fluctuations
as a possible way to reduce these values. Our calculated spin density
matrices resemble those obtained from the molecular--field exchange model 
\cite{Aoki}.

Fig. \ref{LDA+DMFT-AFM}a shows our calculated band structure along major
high symmetry directions of the BZ. A few energy bands are seen to cross the
Fermi level indicating the metallic nature of the solution. The most
striking finding of our AFM calculation is however the existence of the Weyl
points in close proximity to the Fermi level which appear exactly along the $%
\Gamma Z$ line of the BZ, serving here as the magnetization direction. Since
the Weyl points act as Dirac monopoles in k--space, we verify their precise
locations by computing the Berry flux that originates from each point. The
corresponding band structure is shown in Fig. \ref{LDA+DMFT-AFM}b, which
indicates that the Weyl points are of type II according to classification
introduced in Ref. \cite{TypeII}. Fig. \ref{LDA+DMFT-AFM}c shows the
positions of these Weyl points in the Brillouin Zone. Negative
magnetoresistance measurements have been reported for this compound \cite%
{NegR} which can be an indication that the chiral anomaly characteristic for
Weyl semimetals exists here.

To understand the physical origin behind magnetization induced Weyl state in
UNiSn, we introduce a two orbital $k\cdot p$ model for two relativistic
orbitals with an inversion breaking term.

\begin{widetext}
The Hamiltonian reads 
\begin{equation}
H_{eff}=\left( 
\begin{matrix}
A(\mathbf{k})+\Delta  & 0 & Pk_{z}+iVk_{x}k_{y} & Pk_{-}+Vk_{z}k_{+} \\ 
0 & A(\mathbf{k})-\Delta  & Pk_{+}-Vk_{z}k_{-} & -Pk_{z}-iVk_{x}k_{y} \\ 
Pk_{z}-iVk_{x}k_{y} & Pk_{-}-Vk_{z}k_{+} & B(\mathbf{k})+\Delta  & 0 \\ 
Pk_{+}+Vk_{z}k_{-} & -Pk_{z}+iVk_{x}k_{y} & 0 & B(\mathbf{k})-\Delta 
\end{matrix}%
\right)   \label{KP}
\end{equation}%
\end{widetext}where $k_{\pm }=k_{x}\pm ik_{y}$ and bands $A(\mathbf{k}%
)=A_{0}+A_{1}\mathbf{k}^{2}$ , $B(\mathbf{k})=B_{0}+B_{1}\mathbf{k}^{2}$
assume some quadratic (possibly different dispersion). The parameter $P$
controls the inversion breaking. This model was previously used to describe
topological insulator and Weyl semimetal phases in zincblende--like
structures\cite{CuF}. Here we apply a Zeeman splitting to each band A and B
by setting the parameter $\Delta \neq 0$ along the magnetization (z) axis$.$
Once the effective "spin up" and "spin down" states cross, they produce a
Weyl point exactly along 001 direction in the BZ while the gap between these
bands is open for all other k--points provided the inversion breaking $P$
term is non zero. We illustrate this behavior in Fig. \ref{FigKP} which
shows the dispersion of two middle eigenvalues of Eq. \ref{KP}, which are
seen to produce the Weyl point along $\Gamma Z.$ Note also that setting $P=0$
will produce chiral nodal lines.

One of the most striking features of Weyl semimetals is the presence of the
Fermi arcs in their one--electron surface spectra \cite{Arcs}. Although
computations of their shapes are possible via a self--consistent supercell
(slab) calculation of the surface energy bands, given the variety of regular
Fermi states that emerge from our AFM\ calculation together with the fact
that the\ Weyl points are not exactly pinned at the Fermi level, makes it
hard to resolve them in the surface spectrum. Nevertheless, since the arcs
connect the Weyl points of different chirality, one can expect the existence
of long arc--like features in UNiSn that should be protected from
perturbations such as disorder\cite{Disorder}.

In conclusion, based on a computational approach combining density
functional theory of electronic structure and dynamical mean field theory of
strong correlations, we showed that two topological phases of quantum
matter, topological insulator and Weyl semimetal, accompany the
unconventional insulator--metal transition in the 5f electron compound
UNiSn. We uncovered physical origin of its topological insulator behavior
via the occurrence of multiple band inversions between localized
f--electrons and regular band states. We also concluded that the magnetic
ordering triggers the Weyl state with the nodal points appearing along the
magnetization direction. Our study reveals interesting opportunities for
finding other topological phase transitions in strongly correlated systems.
Of particular interest are some non--centrosymmetric actinide compounds
where the interplay between delocalized band electrons and correlated
f--states together with the large spin--orbit coupling is expected to
provide a new playground for studying topological properties of interacting
electrons.

The work was supported by NSF DMR Grant No. 1411336. X. G. Wan was supported
by NSF of China, Grant No. 11525417.

\end{document}